\RequirePackage{graphicx}
\documentclass[aps,pra,floatfix,reprint,longbibliography]{revtex4-1}

\usepackage{times}
\usepackage{graphicx}
\usepackage{amsmath}
\usepackage{amssymb}
\usepackage{enumitem}
\usepackage{mathtools}
\usepackage{xcolor}
\usepackage{float}

\begin{document}	

\title{Efficient Two-Electron Ansatz for Benchmarking Quantum Chemistry on a Quantum Computer}
\author{Scott E. Smart and David A. Mazziotti}
\email{For correspondence, damazz@uchicago.edu}

\affiliation{Department of Chemistry and The James Franck Institute, The University of Chicago, Chicago, IL 60637 USA}

\date{Submitted December 18, 2019; Revised March 13, 2020}

\begin{abstract}
Quantum chemistry provides key applications for near-term quantum computing, but these are greatly complicated by the presence of noise. In this work we present an efficient ansatz for the computation of two-electron atoms and molecules within a hybrid quantum-classical algorithm. The ansatz exploits the fundamental structure of the two-electron system, and treating the nonlocal and local degrees of freedom on the quantum and classical computers, respectively. Here the nonlocal degrees of freedom scale linearly with respect to basis-set size, giving a linear ansatz with only $\mathcal{O}(1)$ circuit preparations required for reduced state tomography. We implement this benchmark with error mitigation on two publicly available quantum computers, calculating accurate dissociation curves for 4- and 6- qubit calculations of ${\rm H}_\textrm{2}^{}$ and ${\rm H}_\textrm{3}^+$.
\end{abstract}
\maketitle

\section{Introduction}

Quantum computers possess a natural affinity for quantum simulation and can transform exponentially scaling problems into polynomial ones~\cite{Abrams1999,Santagati2018,Babbush2014}. Quantum supremacy, the ability of a quantum computer to surpass its classical counterpart on a designated task with lower asymptotic scaling, is potentially realizable for the simulation of quantum many-electron systems~\cite{McArdle2018b,Kassal2010}. Work over the previous decade has been towards this goal with a focus on calculating the energy of small molecules and exploring strategies to leverage emerging quantum technologies, especially those designed to correct or mitigate quantum errors~\cite{Kandala2017,Kandala2019,Moll2017}. In this paper we introduce an efficient ansatz for a two-electron quantum-mechanical system that can be employed as a benchmark for assessing the capabilities and accuracy of quantum computers.  \textcolor{black}{The twin goals of the work are: ({\em i}) to present a quantum-computing benchmark based on the correlated but polynomial scaling two-electron problem, solvable on classical computers, that can be used to assess the accuracy of quantum computers, and ({\em ii}) to develop an efficient ansatz for solving the two-electron problem on quantum computers, based on an effective partitioning of the computational work between classical and quantum computers that is applicable to more general $N$-electron molecular systems.}

The two-electron density matrix (2-DM) of any two-electron system can be expressed as a functional of its one-electron reduced density matrix (1-RDM) and a set of phase factors. This representation of the 2-DM has important connections to natural-orbital functional theories and geminal-based theories in quantum chemistry~\cite{Piris2007, Piris2017a, Coleman2000, Mazziotti2000gem, Mazziotti2001b, Neuscamman2012, Johnson2013, Tecmer2014}.  It offers a natural separation between the nonlocal and local fermionic degrees of freedom in the system~\cite{Walter2016}, scaling linearly and polynomially respectively, and can be leveraged in a variational hybrid quantum-classical algorithm. The entangled, nonlocal degrees are treated on the quantum computer while the local degrees are treated on the classical computer, leading to an efficient simulation of the system.

For a quantum algorithm to exhibit quantum supremacy, obtaining the solution classically will be impractical except for cases that are close to the classical limits of feasibility~\cite{McArdle2018b, Kassal2010}. For some problems such as prime factorization, the solution can be quickly verified, but for many-body quantum systems this is not the case~\cite{McClean2017, Temme2017, Bonet-Monroig2018, Sagastizabal2019, Preskill2018}. Possessing `easy', classically solvable problems to implement and verify will be crucial to evaluate the performance of quantum devices and error mitigation schemes~\cite{Wecker2015}.  Our proposed quantum-classical hybrid algorithm targets only the necessary entanglement needed on the quantum computer, scales linearly with respect to basis size, and has $\mathcal{O}(1)$ circuit preparations, making it an ideal benchmark for molecular simulation. We highlight this by evaluating this ansatz through the computation of ${\rm H_2^{}}$ and ${\rm H}^{+}_{3}$ on two generations of publicly available quantum devices.

\section{Theory}
Because the representation of the 2-DM in terms of the 1-RDM and phase factors has been well studied elsewhere~\cite{Lowdin1956,Lowdin1960,Zumino1962,Coleman2000}, we present in section~IIA only the aspects of the theory that are relevant to the quantum-classical hybrid algorithm in section IIB. We also discuss the preparation of the linear-scaling ansatz in section ~IIC and practical error mitigation techniques in section IID that are employed in the benchmarks in section~III.

\subsection{Structure of the Two-Electron System}
For a two-electron system the energy is given as the trace of the Hamiltonian and the density matrix:
\begin{equation}
E = \text{Tr} ~ (H D)
\end{equation}
where $H$ and $D$ are $2r \times 2r$ with $2r$ being the rank of the one-electron basis set and
\begin{equation}
D^{ij}_{kl} = g_{ij}g^*_{kl}
\end{equation}
in which the wavefunction expansion coefficients $g_{ij}$ are elements of the coefficient matrix $G$. From the antisymmetric nature of fermions, $G$ must be a skew-symmetric matrix, and from a theorem by Zumino~\cite{Zumino1962}, $G$ must have a block-diagonal form $\tilde{G}$ with $2\times 2$ matrices $\mathcal{G}_i$:
\begin{equation}\label{gem}
\tilde{G} = \textrm{diag}~ ({\mathcal{G}}_0,\mathcal{G}_1...\mathcal{G}_r ),
\end{equation}
\begin{equation}\label{gem2}
{\color{black} \mathcal{G}_i }= \begin{pmatrix}
0 & \tilde{g}_{ii'} \\
\tilde{g}_{i'i} & 0  \\
\end{pmatrix}
=
\begin{pmatrix}
0 & \tilde{g}_{ii'} \\
-\tilde{g}_{ii'} & 0  \\
\end{pmatrix}
\end{equation}
The block-diagonal form of $G$ in Zumino's theorem defines an orbital basis set with a natural pairing of the orbitals where we denote the indices of an orbital and its pair by $i$ and $i'$, respectively.

The 2-DM in Zumino's basis has only nonzero elements of the form:
\begin{equation}\label{rdm2no}
\tilde{D}^{ii'}_{kk'} = \tilde{g}^{}_{ii'} \tilde{g}^{*}_{kk'}.
\end{equation}
The 1-RDM, containing the one-body information, can be obtained from the 2-DM by contraction:
\begin{align}\label{rdm1no}
^1 \tilde{D}^i_i = \sum_k {\tilde{D}}^{ik}_{ik} &= \tilde{D}^{ii'}_{ii'}  = \tilde{g}^{}_{ii'}\tilde{g}^*_{ii'}=n_i=n_{i'},\\
\label{rdm1no_b} ^1 \tilde{D}^i_j &= 0,
\end{align}
\textcolor{black}{where all 2-DM elements in the contraction vanish except when $k=i'$.} Because the 1-RDM is diagonal in Zumino's basis set, we find that Zumino's basis set is identically the natural-orbital basis set and that the occupations $n_i$ are the natural orbital occupations. The paired orbitals $i$ and $i'$, we observe, have equal occupations $n_i$ and $n_{i'}$. For a system of two electrons with $S_z=0$, these paired orbitals share the same spatial component with different spin components, denoted by convention as $\alpha$ and $\beta$. This decomposition can be viewed as a particular case ($N=2$) of a more general result derived by Schmidt~\cite{Schmidt1907}, and later by Carlson and Keller~\cite{Carlson1961}. The importance of this decomposition as a quantum computing ansatz for two electrons will be manifest below.

\subsection{Variational Hybrid Algorithm}

Hybrid quantum-classical algorithms with a variational \textcolor{black}{eigenvalue solver} are among the most promising algorithms for near term applications~\cite{Peruzzo2014, McClean2016, Kandala2017, Grimsley2018}. For our approach we utilize a variational eigensolver on the quantum device but apply it only to the optimization of the 2-DM in the natural-orbital basis set. Optimization of the natural orbitals by orbital rotations is performed with polynomial scaling on the classical computer.  \textcolor{black}{In this manner} we are able to partition the nonlocal and local degrees of freedom between the quantum and classical calculations respectively.

Using the structure given in Eqs. \eqref{gem}-\eqref{rdm1no_b}, we see that to evaluate the $\tilde{D}$ matrix, we need only: (1) the natural orbital occupations, measured as the orbital populations on the quantum computer, and (2) the phase corresponding to the natural orbital coefficients, which for a real wavefunction, is simply the parity of the term {\color{black} that can easily be measured on a quantum computer}. That is, we need the phase
$\xi_{ii'}$ where
\begin{equation}\label{phase}
\tilde{g}_{ii'} = \sqrt{n_i}\xi_{ii'}.
\end{equation}
\textcolor{black}{In general, the phases can be measured through tomography on the quantum computer of certain terms of ${\tilde{D}}$ requiring only $\mathcal{O}(1)$ additional circuit preparations. The details of the specific ansatz for the tomography are discussed in the next section.}

After convergence criteria in the optimization of $\tilde{D}$ {\color{black} on the quantum computer} is satisfied through gradient-free optimization (see Appendix A), we optimize the energy on the classical computer {\color{black} through} one-body unitary transformations of the Hamiltonian. Specifically, we optimize the orbitals through a series of Givens rotations.  \textcolor{black}{These complementary quantum and classical optimizations are sequentially repeated until the energy and 2-DM converge.}

\subsection{Preparation of the Efficient Quantum Ansatz}

To create a state of the form in Eq.~\eqref{gem} on the quantum computer, we need to implement double excitations from orbitals $ii'$ to $kk'$. If we consider an initial wavefunction $\tilde{G}_0$ from a standard Hartree-Fock calculation, the ansatz to generate a generic $\tilde{G}$ is:
\begin{equation}\label{exp}
\tilde{G} =  \prod_{i=1}^{r-1} \big(\exp ~{t}_{ii'}^{\bar{i}\bar{i}'}  \big) \tilde{G}_0,
\end{equation}
where $\bar{i}=i+1$, and $t_{ii'}^{\bar{i}\bar{i}'}$ is an antihermitian, antisymmetric two-body matrix with nonzero elements corresponding to an excitation between $i,i'$ and $(i+1),(i+1)'$. The operators acting on $\tilde{G}_0$ can be easily expressed in second quantization as shown in the Appendix. The ansatz is a subset of the unitary coupled cluster (UCC)~\cite{Romero_2018} or antihermitian contracted Schr{\"o}dinger equation (ACSE) ansatz~\cite{Mazziotti2007acse}. From there we perform a Jordan-Wigner transformation (though others may be utilized) which yields an exponential of Pauli strings that are implementable on a quantum device as strings of CNOT gates~\cite{Nielsen2010}. Additionally, the implementation naturally requires only a nearest neighbor connectivity among qubits.

Finally, the tomography of the state involves only the measurement of the orbital occupations for a given qubit \textcolor{black}{in the computational basis as well as the sign. Because there are only $r-1$ phase terms since the last phase is equivalent to a global phase, we only require the tomography of a linear number of sequential terms in $\tilde{D}$ of the form $\tilde{D}^{jj'}_{ii'}$ with $j=i+1$.  Because terms like $\tilde{D}^{jj'}_{ii'}$ and $\tilde{D}^{(j+2)(j+2)'}_{(i+2)(i+2)'}$ are qubit-wise commuting, we can measure $r/2$ terms simultaneously, leading to a constant number of circuit preparations.  In this work we evaluated ${\rm H}_2$ and ${\rm H}_3^{+}$ using the Jordan-Wigner transformation in 4- and 6- qubit cases. There is only one phase term in the 4-qubit calculations of ${\rm H}_2$, which we measured by direct tomography on the quantum computer,  whereas the two phases of the 6-qubit calculation of ${\rm H}_3^+$ were computed by optimization on the classical computer to avoid degradation from noise on the quantum computer.}

\subsection{Error Mitigation Strategies}
Even if we model a two-electron system on a quantum computer and construct the state through the above tomography, we may find that our occupations $n_i$ and $n_{i'}$ do not match for a given $i$, which implies a violation of the fermion statistics. Because the two-electron ansatz in Section IIB formally guarantees a two-electron wavefunction, any deviation in pure-state $N$-representability (up to sampling errors) is due to errors on the quantum computer~\cite{Altunbulak2008,Mazziotti2016b,Coleman2000}.

The effect of errors on current quantum computers can easily influence the $N$-representability of a system~\cite{Coleman2000,Mazziotti2012b} with the extent somewhat depending on the fermionic mapping. For a compact mapping, $S_z$ and $N $ will typically remain constant, but for more general mappings, this is not the case. Other errors can also accumulate, making it difficult to reach certain extrema of the set of density matrices. To address this, we use a projective technique where we map the set of accessible points onto the ideal set of points (see the Appendix). We achieve this by finding an affine transformation $A$ that maps from the accessible but error-prone set $S'$ to the ideal set $S$.

For a general mapping, it is easy for the quantum system to violate $N$ and $S_z$. By utilizing a symmetry verification technique~\cite{Bonet-Monroig2018,Sagastizabal2019}, along with the structure of our tomography requiring only measurements of diagonal terms, we can filter out results which do not obey the correct $N$ and $S_z$ values which effectively projects the resulting state into an eigenstate of the chosen operator. This can be extended to other operators $S$ which commute with the Hamiltonian:
\begin{equation}
[S,H] = 0.
\end{equation}
As will be seen in the results, the symmetry verification is useful in bringing the results back to the set of all two-electron states, and then the projection restores the equality of the two pairing-related sets of occupations, ${n_i}$ and $n_{i'}$.

\section{Results}
Using the two-electron ansatz, we first treat the molecular dissociation of ${\rm H}_2$ in a minimal {\color{black} Slater-type-orbital-expanded-in-three-Gaussians (STO-3G)} basis set of two electrons in four orbitals. The quantum algorithm is implemented on both the 5- and 14- qubit devices, denoted as ibm-5 and ibm-14, representing two generations of superconducting quantum devices by IBM. With the Jordan-Wigner transformation~\cite{Jordan1928} the system can be represented with 4-qubits {\color{black} though more compact mappings are certainly possible. Note in this basis only a single excitation is possible.} Figure 1 shows the potential energy curve of the ${\rm H}_2$ molecule, computed with full error mitigation.

\begin{figure}
\begin{center}
\includegraphics[scale=0.57]{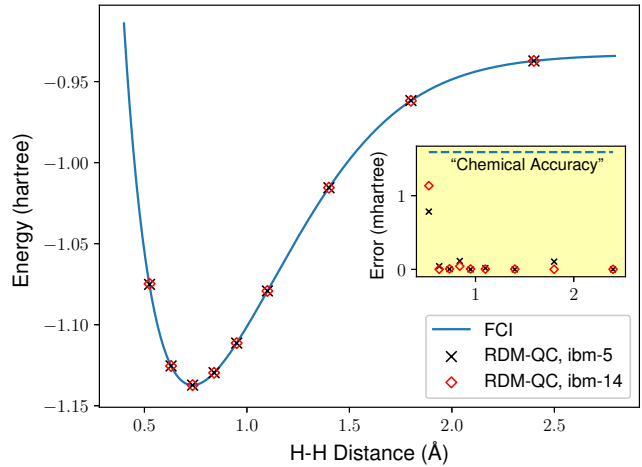}
\caption{Dissociation curves for the ground state of ${\rm H}_2$ from the variational quantum algorithm on the quantum computer and the full configuration interaction are shown. Both results were run with 4 qubits, but on 5- and 14- qubit frameworks. The inset shows the difference in energy from the FCI results in mhartrees. The increased error for the shortest distance relates to the difficulty in reaching the Hartree-Fock state on a quantum computer when using entangling gates. For more experimental details, see the Appendix.}
\end{center}
\end{figure}

With the help of error mitigation techniques, both quantum devices are able to capture the dissociation of the molecule, achieving mhartree accuracy across the spectrum of states, leaving differences in the devices somewhat unclear. Inspection of the device calibration (Appendix A, Table II) indicate that large measurement errors likely occur on ibm-5 with superior performance expected from ibm-14. {\color{black} Scans of the 1-RDMs with respect to the $t_{11^\prime}^{22^\prime}$ parameter controlling the single double excitation are shown in Figure 2. The ibm-5 device is seen on the top row, and the ibm-14 on the bottom row, and we also show the effect of the symmetry verification in correcting the occupations.}

\begin{figure}
\begin{center}
\includegraphics[scale=0.70]{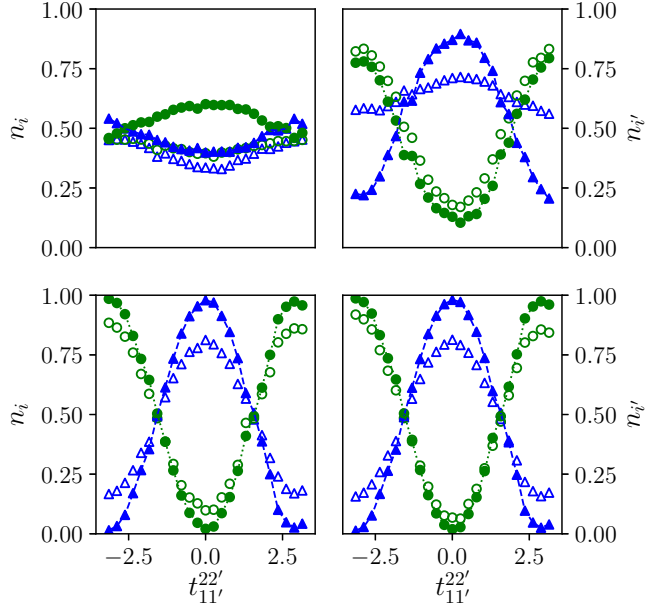}
\caption{Measured (unordered) occupation numbers with respect to the {\color{black} coefficient $t^{22^\prime}_{11^\prime}$ of the single double excitation operator} within the entangling circuit. Blue triangles represent the first qubit, and green circles represent the second qubit, unordered. Unfilled objects show the raw occupations, and filled objects the symmetry-corrected occupations. Uncertainties for the values are similar in size to the markers. }
\end{center}
\end{figure}

While the ibm-5 device maintains continuity with respect to $t^{22^\prime}_{11^\prime}$, it has distinct problems. First, we observe {\color{black} (see top left insert)} correlated measurement error in the set of qubit occupation $\{ n_i \}$ which cause an inversion in the expected relationship between $n_1$ and $n_2$ among the $i$ ($\alpha$) occupations. Second, we find that symmetry verification is effective in increasing the differentiability of the two states (note for a decohered system, $n_1$ and $n_2$ would be identically 0.5), yet it is not able to correct for the reversal in ordering seen on the qubits, and somewhat reinforces this error for $n_2$.

On ibm-14, while there is still a contraction of the occupation numbers across the range of $t^{22^\prime}_{11^\prime}$, the obtained curve is continuous, and the two sets of occupations correspond to expected values. With symmetry verification, we obtain nearly the ideal occupations, which otherwise are far from spanning the full spectrum of occupations. The effect of our further correction is to stretch the sinusoidal curves in Fig.~2 so that their maxima and minima are $1.0$ and $0.0$ respectively, hence it is not shown here.

To quantify the effects of symmetry verification, we calculate the area between the two changing orbital occupations for both $\{n_i\}$ and $\{n_{i'} \}$ (denoted as $V_i$ and $V_{i'}$, {\color{black} where $V_i = n_2-n_1$, and $V_{i'} = n_{2'}-n_{1'}$}, ) subject to different symmetries, as well as the uncertainty in measurement after each symmetry is applied, and show these in Table 1. The maximum and minimum values for this metric would be $2$ and $0$, representing fully error-free and fully decohered states respectively. In each case, application of multiple symmetries serves to increase the resulting `reach' of the state, without increasing variance with respect to decreased measurement counts.

\begin{table}
\caption{The table shows the area between the two occupations ($V$) obtained for the entangling circuit described in the text for the two different quantum devices. The $V_i$ and $V_{i'}$ values were calculated separately. Here we have the 5- and 14-qubit devices, respectively, given in a 95\% confidence interval due to sampling. The left column indicates the values with the different applied symmetries.}
\begin{center}
\begin{tabular}{c|cc|cc}
Device & \multicolumn{2}{c|}{5-qubit} & \multicolumn{2}{c}{14-qubit} \\
& ~~~ V$_i$ ~~~& ~~~V$_{i'}$~~~ & ~~~ V$_i$ ~~~& ~~~V$_{i'}$~~~ \\
\hline
None    & 0.096$\pm$0.005 & 0.861$\pm$0.007  & 1.405$\pm$0.007 & 1.476$\pm$0.007 \\
${N}$     & 0.18$\pm$0.01 & 0.75$\pm$0.01 & 1.89$\pm$0.01 & 1.907$\pm$0.009  \\
${S}_z$   & 0.25$\pm$0.01 & 1.06$\pm$0.02 & 1.723$\pm$0.009 & 1.736$\pm$0.009  \\
${N},{S}_z$ & 0.33$\pm$0.02 & 1.43$\pm$0.02 & 1.93$\pm$0.01 & 1.94$\pm$0.01
\end{tabular}
\end{center}
\end{table}

%
Computations with 6 orbitals are performed only on the ibm-14 device since more than 5 qubits are required. Using simplifications seen in Nam et al.~\cite{Nam2019}, we are able to construct a gate with 8 CNOT gates which still requires only neighboring connections (see Appendix C). Due to the longer depth of the circuit, the effects of noise are more pronounced and we find it difficult to reliably measure the phase of the 2-DM terms required in Eq.~\eqref{phase} and note that these are not always continuous. To show the overall effect of errors on the 6-qubit system on the local occupations, we present a scan of possible symmetry verified 1-RDMs over a range of the parametrized entangling gates in Fig.~3. Additional details regarding the computation are provided in the Appendix. While we do not show the occupations in terms of the parameters, the effect of the aggregate errors for this case is again to shrink the portion of the hyperplane accessible to the quantum device.

\begin{figure}\label{h3nrep}
\begin{center}
\includegraphics[scale=0.28]{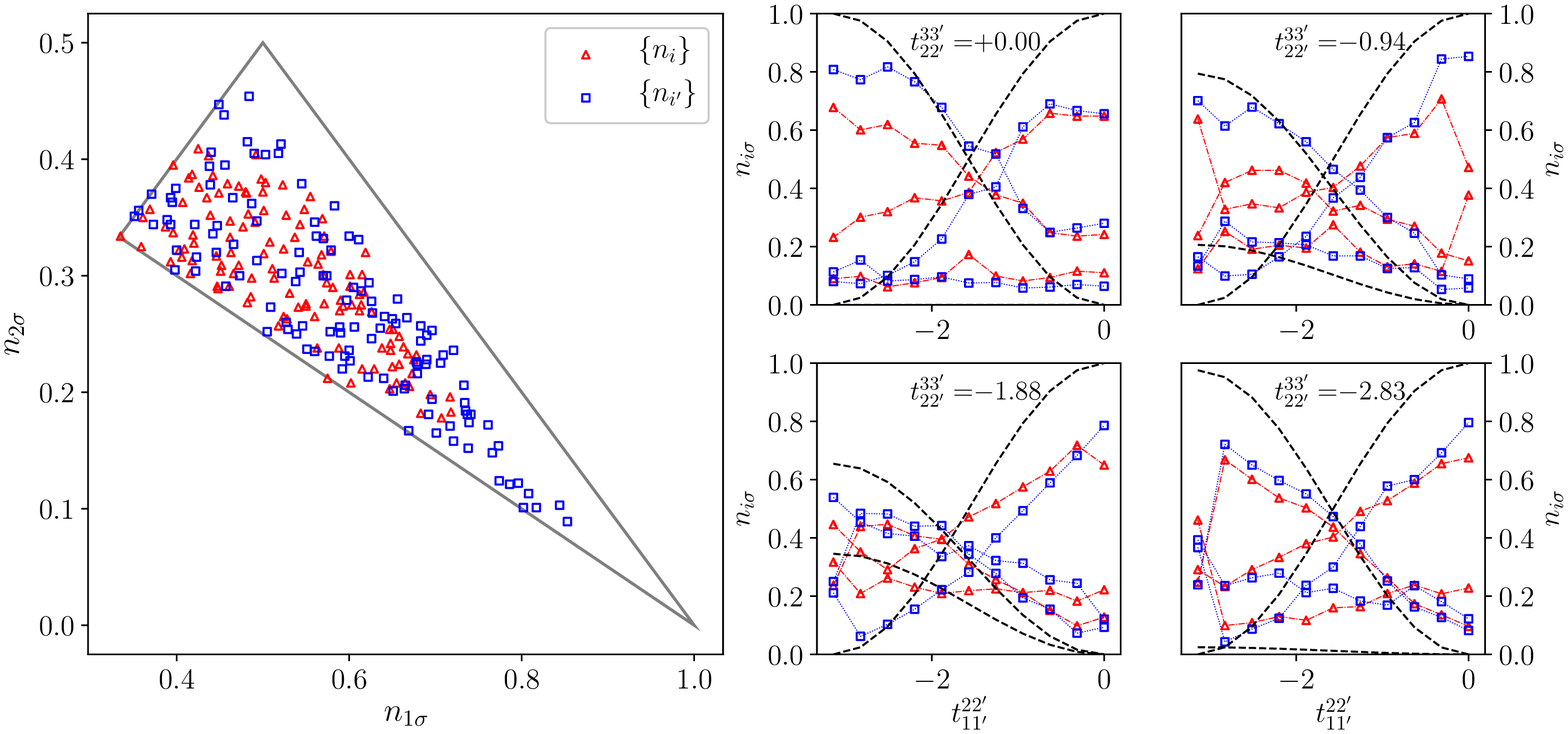}
\caption{{\color{black}(Left) measured occupations of the 6-qubit system following symmetry application of $N$ and $S_z$, as well as the boundaries of the ideal polytope. Note, $\sigma$ refers to the use of $i$ or $i'$. Because these occupations correspond with an $N$-representable system, the equality $n_{1\sigma}+n_{2\sigma}+n_{3\sigma}=1$ holds, and so we show the orthographic projection along the $n_{3\sigma}$ axis. The device was sampled over the range $t^{22^\prime}_{11^\prime}, t^{33^\prime}_{22^\prime} \in  [-\pi,0]$ in $\frac{\pi}{10}$ intervals for the entangling parameters. The ratio of areas of the experimental and theoretical convex hulls of these obtained points is $0.48$ and $0.68$ for $\{n_i\}$ and $\{ n_{i'}\}$, respectively. (Right) Unordered $n_{1\sigma}$, $n_{2 \sigma}$, and $n_{3 \sigma}$ occupations with respect to parameters $t^{22^\prime}_{11^\prime}$ at different slices of $t^{33^\prime}_{22^\prime}$ (given in each plot). Dashed lines indicate the ideal occupations for the unordered qubits (equivalent for $n_i,n_{i\prime}$). The set of occupations closer to $(n_1,n_2,n_3)=(\frac{1}{3},\frac{1}{3},\frac{1}{3})$ is well covered, though there is clear difficulty in reaching uncorrelated regions, where only one occupation is 1.}}
\end{center}
\end{figure}

Expectedly, the obtained results differ greatly depending on the qubits and available connectivity of the quantum device, though in general we still observe a degree of continuity in the local 1-RDM properties. By dealing with the phases classically, we are able to calculate the dissociation curve for triangular ${\rm H}_3^+$ in Fig.~4. Again, we are able to obtain chemically accurate energies across the dissociation curve, although there was difficulty in sampling the uncorrelated Hartree-Fock state which is a vertex of the polytope. The error mitigation techniques we use also extend the capabilities of noise-limited quantum computers, which otherwise do not span the ideal $N$-representability of the state~\cite{Smart2019b}.

\begin{figure}[H]
\begin{center}
\includegraphics[scale=0.56]{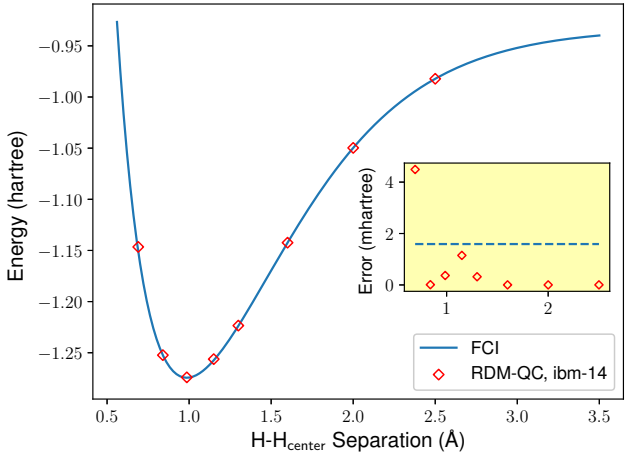}
\caption{Dissociation curves for the ground state of ${\rm H}_3^+$ from the variational quantum algorithm on the quantum computer and the full configuration interaction. The experiment was run on a 14- qubit framework. The inset shows the difference in energy from the FCI results in mhartrees. The increased error for the shortest distance relates to the difficulty in reaching the Hartree-Fock state on a quantum computer when using entangling gates. For more experimental details, see the Appendix.}
\end{center}
\end{figure}

\section{Discussion and Conclusion}
In this work we present an ansatz for two-electron quantum systems which can be implemented on near-term and future quantum computers. Applying this on two public-access quantum computers highlights the successes and differences of two generations of quantum computers, as well as the difficulties which must be overcome in approaching more complicated systems. We also show that using error mitigation strategies we are able to simulate both ${\rm H}_2$ and ${\rm H}_3^+$ to high accuracy. The proposed ansatz can be readily applied to two-electron atoms and molecules in larger basis sets with similar types of error mitigation. The gate sequence proposed here can be applied as a generic ansatz, removing the need for long expansions of the required exponential operators.

\textcolor{black}{The ansatz is efficient mainly in regards to its scalability with respect to other methods of state preparation. Unitary coupled cluster, which for a two-electron system only needs to be expressed with single and double excitations (i.e., UCCSD), and which additionally has only one occupied orbital for the $\alpha$ and $\beta$ excitations, still will have $\mathcal{O}(r^2)$ terms in the ansatz. Furthermore, quantum tomography would additionally require the measurement of the $\alpha \beta$ block of the 2-DM, which naively has $\mathcal{O}(r^4)$ terms, and hence, the method would scale as $\mathcal{O}( r^{6} )$ where the depth $d$ and the number of measurements contribute factors of $r^{2}$ and $r^{4}$, respectively.  Other methods based on the propagation of the Hamiltonian could be implemented, but they also have large costs and would not necessarily lead to the same advantages that result from the structure of the 2-DM.}

From the results, it is clear one cannot rely solely on the energy or other external molecular properties to investigate the integrity of the quantum device, particularly in comparing the performance of the two-qubit devices on the 4-qubit calculation. While averaged or localized metrics related to qubit depolarization, dephasing, or bit-flip errors are often used as indicators of performance of the quantum device, they may not translate directly to the fidelity of a simulated fermionic system, particularly with multi-qubit or environmental effects. Looking at these metrics in conjunction with the physical properties of benchmark problems like the ground-state energy of two-electron atoms and molecules will yield greater insights into the fidelity of a quantum device and its needs for error mitigation or correction.

The method of symmetry verification for error mitigation is useful in that its different forms are low cost and can easily correct flagrant faults in the output, such as particle count (or more generally, parity) and the projected spin symmetries. Others forms of error mitigation related to the reduced density matrices (RDMs) can be implemented as constraints on the tomography from $N$-representability, or in a form of post-correction of the two-electron RDM (2-RDM) where the measured 2-RDM is purified through semidefinite programming, which can be applied to arbitrary $N$-electron systems~\cite{Smart2019b}. The mapping we use can have difficulties when errors begin to change the ordering of occupations for larger and larger systems.

The electron pair itself plays a key role in such phenomena as superconductivity {\color{black} and} bonding, and yet the exact energy of a two-electron system itself cannot be solved exactly with known methods. Such a problem shows the essence of the electron-electron interaction as well as some of the complexities of electron correlation and quantum mechanics. The theory in this work could also be seen as a subset of more complex geminal-based wavefunction methods, which appear in classical electronic structure theory where the electron pair is treated as the fundamental unit to improve the accuracy beyond the mean-field approximation~\cite{Piris2007, Piris2017a, Coleman2000, Mazziotti2000gem, Mazziotti2001b, Neuscamman2012, Johnson2013, Tecmer2014, Boguslawski2014, Stein2014, Head-Marsden2017}. {\color{black} The exploitation of the structure of the 2-DM in the natural-orbital basis set can be extended to more general pairing 2-RDMs for efficient implementations of pairing (geminal) theories or natural orbital functional theories on quantum computers.}

Here we show that the properties of the two-electron system lead to an ansatz which is well suited for use in a hybrid quantum-classical approach, where degrees of freedom that would increase exponentially with $N$ are treated on the quantum computer, while non-exponentially increasing degrees of freedom are treated on the classical computer.  \textcolor{black}{The treatment of the orbital rotations on the classical computer can be generalized to $N$-electron molecular systems where the orbital rotations can be used to implement active-space methods where orbital rotations are used to optimize the correlation in a modest subset of total orbitals known as the active orbitals.}  Generally, such \textcolor{black}{orbital-rotation algorithms} including \textcolor{black}{ active-space self-consistent-field algorithms}~\cite{Roos1980,Schlimgen2016} could assist in achieving quantum supremacy by further lowering resource requirements on the quantum computer as well as tomography and measurement costs. The two-electron system also is useful in its own right as a benchmark for molecular simulation on a quantum computer, where it serves as a simple yet effective way to assess the performance of an arbitrary quantum device. The present work is clearly applicable for the current state of noisy quantum computers, but will continue to be relevant as improving generations of quantum computers are developed. The two-electron ansatz, albeit polynomially scaling even on a classical computer, can serve as a powerful benchmark for quantum computers due to the availability of accurate results from classical computers and the requirements shared by its solution and the solution of exponentially scaling many-electron problems.

\section{Acknowledgements}
The authors acknowledge use of the IBM Q for this work. The views expressed are of the authors and do not reflect the official policy or position of IBM or the IBM Q team. D.A.M. gratefully acknowledges the Department of Energy, Office of Basic Energy Sciences, Grant DE-SC0019215, the U.S. National Science Foundation Grant CHE-1565638, and the U.S. Army Research Office (ARO) Grant W911NF-16-1-0152. {\color{black} We also are thankful to the two reviewers who gave invaluable feedback.}

\section*{Data Availability}

Data is available from the corresponding author upon reasonable request.

\section*{Author Contributions}
D.A.M. conceived of the research project. S.E.S. and D.A.M. developed the theory.  S.E.S. performed the calculations. S.E.S. and D.A.M. discussed the data and wrote the manuscript.

\appendix
\section{Computational Details}
The electronic structure package PySCF ~\cite{Sun2017} was used to obtain the one- and two- electron integrals and to perform restricted Hartree-Fock and full configuration interaction (FCI) calculations.

For the quantum computation we used the IBM Quantum Experience devices Yorktown (Sparrow, 5-qubits) and Melbourne (Albatross, 14-qubits), available online. The former has triangular-type coupling between qubits, and the latter has square-type coupling between qubits. These cloud accessible quantum devices are fixed-frequency transmon qubits with co-planer waveguide resonators~\cite{Koch2007,Chow2011}. The quantum information software development kit QISKIT was used to interface with the device. We include the calibration data in Table~II.

For the 6-qubit case, $2^{11}$ measurements were obtained, and we used a simple Nelder-Mead simplex method with the Han initial simplex for the 6-qubit case on the quantum computer. Classically orbital rotations were performed with Givens rotations, with the Broyden-Fletcher-Goldfarb-Shanno (BFGS) algorithm being utilized. Convergence criteria were more strict for the 4-qubit case, with convergence between the $^2D$ and $^2K$ steps being 1 mH. In the 4- and 6-qubit cases, we chose the best of 2 runs as the optimal results. While this did not make a difference for most points, for those it did, the difference in energies for the 2 runs were usually significant, indicating that noise had led the optimization into some local minima.

%

\begin{table}\caption{Calibration data for the ibm-5 and ibm-14 devices during benchmarking. U$_2$ and U$_3$ represent the errors for single qubit unitaries containing one and two $X_{\pi /2}$ pulses and two and three frame changes respectively. RO represents the readout error, and we have the standard $T_1$ depolarization and $T_2$ dephasing times. [$j$] specifies the target qubit with control qubit $i$, and we report the error. }
\begin{tabular}{c|cc|c|cc|ccc}

\textbf{Qubit} & \textbf{U}$_2$  & \textbf{U}$_3$  & \textbf{RO} & \textbf{T}$_1$  & \textbf{T}$_2$  & \multicolumn{3}{c}{$\left[ j \right]$ \textbf{CX}$_i^j$ } \\
$i$ & ($10^{-3}$) & ($10^{-3}$) &  ($10^{-2}$) & ($\mu s$) & ($\mu s$)  & \multicolumn{3}{c}{($10^{-2}$)}  \\
\hline
\multicolumn{9}{c}{\textbf{ibm-5}} \\
\hline
0                                    & 2.7               & 5.5               & 5               & 46              & 53              & [1] 5.1                                     & [2] 4.2 &     \\
1                                    & 2.9               & 5.8               & 25              & 62              & 53              & [2] 6.8                                     &     &     \\
2                                    & 6.4               & 12.9              & 1               & 85              & 74              &                                         &     &     \\
3                                    & 3.8               & 7.6               & 17              & 61              & 28              & [2] 7.9                                     & [4] 4.0 &     \\
4                                    & 2.8               & 5.7               & 36              & 68              & 62              & [2] 4.2                                     &     &     \\
\hline
\multicolumn{9}{c}{\textbf{ibm-14}} \\
\hline
0                                    & 2.3               & 4.7               & 3               & 62              & 22              &                                         &     &     \\
1                                    & 5.1               & 10.1              & 10              & 54              & 101             & [0] 3.7                                     & [2] 6.4 &     \\
2                                    & 3.9               & 7.8               & 5               & 75              & 168             & [3] 6.7                                     &     &     \\
3                                    & 1.5               & 3.0               & 27              & 63              & 51              &                                         &     &     \\
4                                    & 2.4               & 4.8               & 6               & 56              & 34              & [3] 5.6                                     & [10] 5.4 &     \\
5                                    & 2.3               & 4.6               & 4               & 24              & 46              & [4] 6.1                                     & [6] 7.5 & [9] 5.9 \\
6                                    & 2.3               & 4.6               & 4               & 77              & 53              & [8] 2.9                                     &     &     \\
7                                    & 1.3               & 2.7               & 16              & 50              & 82              & [8] 2.3                                     &     &     \\
8                                    & 1.5               & 3.0               & 4               & 125             & 183             &                                         &     &     \\
9                                    & 2.8               & 5.6               & 4               & 44              & 65              & [8] 7.0                                     & [10] 4.0 &     \\
10                                   & 2.5               & 5.0               & 4               & 51              & 55              &                                         &     &     \\
11                                   & 181               & 362               & 34              & 63              & 102             & [3] 14                                      & [10] 10  & [12] 11  \\
12                                   & 3.7               & 7.3               & 9               & 89              & 177             & [2] 7.3                                     &     &     \\
13                                   & 5.1               & 10.3              & 4               & 26              & 59              & [1] 13                                      & [12] 3.9 &
\end{tabular}
\end{table}

\section{Error Mitigation with N-Representability of $\wedge^2 \mathcal{H}_n$}
The error correction is similar to previous work where we look for a transformation $A$ to map the experimental polytope $S'$ to the correct polytope $S$. The structure of the N-representability conditions is such that the ordering inequalities applied to each of the half-sets ($\{ n_i\}$ and $\{ n_{i'}\}$)  describe a hyperplane. The vertices $V^{r}$ can be described as a set with elements $v^{r}_j$ ($r$ spatial orbitals):
\begin{equation}
v^{r}_j = [\frac{H(j-1)}{j},\frac{H(j-2)}{j},...,\frac{H(j-r)}{j}]
\end{equation}
where $1\leq j\leq r$ and $H(x)$ is the Heaviside step function. The vertices for $\wedge^2 \mathcal{H}_4$ are given as $v_1^2 = (1,0)$, $v_2^2 = (\frac{1}{2},\frac{1}{2})$. The vertices for $\wedge^2 \mathcal{H}_6$ are then: $v_1^3 = (1,0,0)$, $v_2^3 = (\frac{1}{2},\frac{1}{2},0)$, $v_3^3 = (\frac{1}{3},\frac{1}{3},\frac{1}{3})$. These form a $r-1$ dimensional hyperplane in the $r$ dimensional subspace for the two half sets, respectively. The practical effect of symmetry verification here (mostly from the $S_z$ application) is to project noisy points onto the plane. The effect of the N-representability application then is to map the measured points to the extreme points of $V^{n'}$. This can be visualized in Fig. \eqref{h3nrep} by mapping the accessible triangular plane to the black outlined plane.

\begin{figure}[th]
\begin{center}
\includegraphics[scale=0.70]{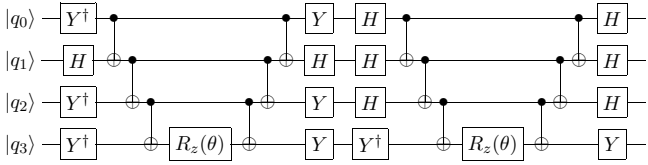}
\caption{The unitary coupled cluster term for two Pauli terms, here representing the exponent of $Y_0 X_1 Y_2 Y_3$ followed by $X_0 X_1 X_2 Y_3$. For optimal connectivity (i.e., no intermediate CNOT gates for intermediate orbitals), there are 12 CNOT operations, generally the most error prone step.}
\end{center}
\end{figure}

\begin{figure}[th]
\begin{center}
\includegraphics[scale=0.75]{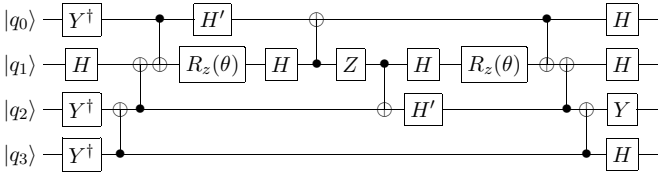}
\caption{The unitary coupled cluster term as seen in Fig. 5, but in a simplified form. Note the target qubit for the rotation here is not $q_3$ but $q_1$. Importantly, the circuit require only 8 CNOT operations, and still has the same connectivity requirements. The gate $H'$ is defined as $S^\dagger H S$ (applied left to right).}
\end{center}
\end{figure}

For points close to the edges, one can imagine that with significant non-coherent error the projected points might lie outside of the space. We account for this by re-projecting these points into the polytope according to the closest edge. A simple semi-definite program would also suffice, though we take a more geometric approach. For mild errors we find that this method is satisfactory.
\section{Second-Quantization Treatment of Entangling Gates and Phase}
The exponential operator in \eqref{exp} has a readily recognizable form in second quantization. Utilizing creation and annihilation operators in the natural orbital basis the operator:
\begin{equation}
\hat{T}^{i}_{j} = t^{ii'}_{jj'}\hat{a}^\dagger_{i}\hat{a}^\dagger_{i'}\hat{a}^{}_{j'}\hat{a}^{}_{j}
\end{equation}
where $t^{ii'}_{jj'}$ is a scalar, and anti symmetric with respect to swapping lower or upper indices, can be used to construct a two-body unitary operator $U$:
\begin{equation}
U = \exp (\hat{T}^{i}_{j} - \hat{T}^{i\dagger}_{j}).
\end{equation}
Only one exponential Pauli term is needed to excite the initial double excitation. After that, a simplification was performed akin to that of Nam et al.\cite{Nam2019}. For a system of two electrons, because do not consider number or spin changing operations, using the inverse Jordan-Wigner mapping we can simplify the total number of Pauli terms needed in the exponential to two.

To measure the phase of the terms in Eq.~\eqref{phase}, we are interested in tomography of the 2-DM elements. Using the Jordan-Wigner transformation we can approximate the total operator as:
\begin{align}\label{C1}
\langle M_{ij}\rangle &= \langle \hat{a}^\dagger_{i}\hat{a}^\dagger_{i'}\hat{a}^{}_{j'}\hat{a}^{}_{j}+\hat{a}^\dagger_{j}\hat{a}^\dagger_{j'}\hat{a}^{}_{i'}\hat{a}^{}_{i}\rangle \\  &\approx \frac{1}{4} \langle X_i X_j X_{i'} X_{j'} \rangle + \frac{1}{4}\langle X_i X_j Y_{i'} Y_{j'} \rangle \\ \label{C2}
 &\approx \frac{1}{4} \langle X_i X_j X_{i'} X_{j'} \rangle + \frac{1}{4}\langle Y_i Y_j X_{i'} X_{j'} \rangle.
\end{align}
Where we utilize the fact that any non-number conserving elements will be contribute 0. Due to the limited amount of sign terms we need to measure, and the fact every other {\color{black} excitation in the linear sequence will be completely commuting, }we can prepare one circuit with only $X_i$ terms, and another with alternating $X$ and $Y$ pairs which give sign terms of either \eqref{C1} or \eqref{C2}, yielding only 2 circuit preparations and a $\mathcal{O}(1)$ complexity. Note, obtaining all of the proper terms also scales as $\mathcal{O}(1)$, taking no more than 8 additional circuits. {\color{black} This method, however, did not yield significant increases in the accuracy in our computations, and so for our optimizations we used the approximate circuit.}

One issue with a direct measurement of the sign is that while the diagonal elements we measure can be symmetry verified, the $N$ and $S_z$ operators do not necessarily commute with the Pauli terms (despite commuting with the operator as a whole).
$[\hat{N},X_1X_2Y_3Y_4]\neq 0 $.
To partially address this, we attempted to use in-line symmetry measurements, where we cast the Pauli measurement onto an ancilla qubit, allowing by propagating through the circuit to the corresponding entangling gate. This requires explicit connectivity requirements and attention to the layout, as a change in Pauli basis between the applied entangler and the required Pauli term. One issue we found was that while the sign information obtained in this manner was coherent and exhibited the proper behavior and change in sign with respect to $t^{22^\prime}_{11^\prime}$, there was a phase difference on the ancilla which resulted in the sign information being shifted from the magnitude of the occupations. A direct measurement of the Pauli terms, while not allowing for symmetry verification, still contained the correct qualitative information, and so was utilized for the 4-qubit case. For the 6-qubit case, obtaining reliable information throughout the longer optimization requirements was more difficult, and so to ease the demands on the quantum computer, we mapped the sign of the elements to the sign of the ideal function generated by our entanglers (a simple product of sine and cosine functions).

\bibliography{qc_chem,QC3}

\bibliographystyle{apsrev4-1}

\end{document}